# Electrical conductivity of silver halide - cadmium halide systems

Efthimios S. Skordas

*Section of Solid State Physics and Solid Earth Physics Institute, Department of Physics, National and Kapodistrian University of Athens, Panepistimiopolis, Zografos 157 84, Athens, Greece.*

**Abstract**

Very recent measurements of the electrical conductivity of solid systems AgX – $CdX_2$ (where X≡Cl,Br) that form large areas of solid solutions, have shown that maximum conductivity occurs for a concentration around 20 mol% of the cadmium halide. Here, we suggest a quantitative explanation of this phenomenon based on a model that was suggested (J. Appl. Phys. 103, 083552, (2008)) for estimating the compressibility of multiphased mixed crystals. In addition, explicit conditions are obtained which predict when such a conductivity maximum is expected to occur.

# 1. Introduction

The elastic properties of ionic solids provide valuable information about cohesive energies, interatomic forces, and anharmonic properties such as thermal expansion. They are also needed for comparisons with atomic model calculations of elastic constants and for use in lattice and defect property calculations. The alkali halides are the simplest ionic solids and studies of their elastic properties have yielded useful information about interatomic potentials and forces. Their description in terms of the well known Born model is very good. On the other hand, the silver halides exhibit unusual properties compared to the alkali halides, such as appreciably lower melting point and ionic conductivity larger by several orders of magnitude. This is one of the basic reasons that the study of physical properties of silver halides have recently attracted an intense interest. These studies include: First, the recent measurements of the temperature dependence of their elastic constants by means of the resonant ultrasound spectrospopy method [1], which overcomes the difficulty in determining elastic constants of small samples by pulse echo methods because of the short travelling time of pulse echoes. Second, first principles calculations [2] of the structural, electronic and thermodynamic properties of silver halides as well as of their rock-salt $AgCl_xBr_{1-x}$ alloys by application of the full potential linearized augmented plane wave method. For the alloys, the effect of composition on lattice constants, bulk modulus, cohesive energy, bond ionicity, band gap and effective mass was also investigated. Third, the phonons and elastic constants AgCl and AgBr under pressure have been extensively studied [3] by using the pseudopotential plane-wave method within density functional theory. Finally, a state of the art density functional theory study of the intrinsic defects of silver halides has been recently reported [4, 5]. The present manuscript deals also with the study of the intrinsic defects in silver halides in the sense that will be described below.

The structure of the intrinsic point defects in the solid silver halides, and in particular AgCl and AgBr, is of both fundamental and industrial relevance. Due to their unique defect properties, the silver halides are of major importance to the photographic industry. While the majority of rocksalt-structured materials contain Schottky defect pairs, the dominant form in the silver halides are Frenkel defect pairs, consisting of interstitial cations and corresponding vacancies (e.g., see Ref. [6] and references therein). It is the ease of formation of the interstitial cation together with its subsequent high mobility within the crystalline lattice, which are crucial to the efficiency of the photographic process.

It is a well known experimental fact that in AgCl and AgBr their electrical conductivity plot $\ln(\sigma T)$ versus $1/T$ exhibits a strong upwards curvature in the high temperature (T) range reaching before melting conductivity values up to above 10 and 50 $\Omega^{-1}m^{-1}$, respectively [7-9]. This effect has been attributed mainly to an excessive increase of the concentration of cation Frenkel defects near the melting point due to the non-linear decrease of the defect Gibbs formation energy [9] upon increasing the temperature, which is accompanied by a simultaneous *increase* [10] of the defect formation enthalpy and entropy versus the temperature (cf. This entropy differs essentially from the dynamic entropy defined recently in natural time [11]).

In a recent work, Górniak et al. [12] presented conductivity measurements as a function of temperature and composition on solidified mixtures of AgCl or AgBr with respective chloride or bromide of divalent metal. Chief among these mixtures were AgCl-CdCl$_2$ and AgBr-CdBr$_2$ their phase equilibria of which exhibit the following significant feature [13-15]: Very large areas of solid solutions based on the corresponding silver halide are formed. These solid solutions showed an important increase in conductivity with the concentration of Cd$^{2+}$, until around 20 mol% CdX$_2$ (X=Cl, Br). At the maximum, the value of electrical conductivity in the AgCl+ CdCl$_2$ solid solution is about 40 times higher than in pure AgCl and the value of electrical conductivity in AgBr+CdBr$_2$ solid solution is about 3 times higher than in pure AgBr. This increase of conduction was qualitatively discussed by Górniak et al. [12] by the increase in the concentration of vacancies on the cationic sublattice arisen from the addition of $Cd^{2+}$ (since it is well known [16-18] that the addition of aliovalent cations with higher valence than $Ag^+$ produce –for reasons of charge compensation- additional extrinsic cation vacancies). Here, we attempt a quantitative explanation of this phenomenon, i.e., the appearance of the maximum conductivity of the solid solution at a certain concentration of $Cd^{2+}$, by employing an established thermodynamical model –termed $cB\Omega$ model [19-23] (see below). Note that the same model has been employed [24] for the estimation of the compressibility of the mixed alkali halide crystals and led to values in good agreement with detailed experimental data [25-27].

**2. Determination of the composition of the maximum conductivity on the basis of a thermodynamical model.**

If $V_I$ and $V_{II}$ denote the molar volumes of the two pure constituents (*I*) and (*II*), then the "molar" volume V of the solid solution can be written as:

$$V = V_I(1-x) + V_{II} x \qquad (1)$$

where *x* stands for the molar concentration of the crystal *II* in the mixed system. Note that Eq.(1) differs from the usual Vegard's rule (stating that the mean lattice constant of the mixed system is a linear combination of the lattice constants of the end numbers) and has been found to describe well the experimental data in a number of cases [6]. Differentiating Eq.(1) in respect to pressure, we find [6]:

$$\frac{B}{B_I} = \frac{1 + x \dfrac{Nv^d}{V_I}}{1 + x \dfrac{\kappa^d}{\kappa_I} \dfrac{Nv^d}{V_I}} \qquad (2)$$

where $B$ and $B_I$ are the bulk modulus of the mixed crystal and the pure crystal $I$, respectively, and $\kappa_I = 1/B_I$ where $\kappa_I$ is the compressibility of the pure crystal $I$. The quantity $v^d$ is the so called defect volume and represents the difference of the volume of a crystal of $N$ molecules of type $I$ and the same crystal in which one of its "molecules" has been exchanged by a molecule of type $II$, thus -after assuming for the sake of convenience that $V_{II} > V_I$ - it is given by [6]:

$$Nv^d = V_{II} - V_I \qquad (3)$$

Finally, the quantity $\kappa^d$ in Eq.(2) denotes the compressibility of the defect volume $v^d$ defined as

$$\kappa^d = -\frac{1}{v^d} \frac{dv^d}{dP}\bigg|_T \qquad (4)$$

which in general differs from the compressibility $\kappa_I$ of the component $I$, thus

$$\frac{\kappa^d}{\kappa_I}(\equiv \mu) \neq 1 \qquad (5)$$

Combining Eqs.(2), (3) and (5) we find

$$\frac{B}{B_I} = \frac{1+x\lambda}{1+x\mu\lambda} \qquad (6)$$

where

$$\lambda \equiv \frac{V_{II}}{V_I} - 1 > 0 \qquad (7)$$

Hence, multiplying Eqs.(1) and (6) we get:

$$\frac{BV}{B_I V_I} = \frac{(1+x\lambda)^2}{1+x\mu\lambda} \qquad \text{or}$$

$$\frac{B\Omega}{B_I\Omega_I} = \frac{(1+x\lambda)^2}{1+x\mu\lambda} \qquad (8)$$

where $\Omega$ and $\Omega_I$ stand for the mean volume per atom of the mixed system and the "pure" crystal *I*, respectively.

We now turn to the thermodynamical model –termed cBΩ model as mentioned - which states that the defect Gibbs activation energy $g^{act}$ is interconnected with the bulk quantities through the relation:

$$g^{act} = c^{act} B\Omega \qquad (9)$$

where $c^{act}$ is independent of temperature and pressure. The validity of this relation has been carefully checked in a variety of solids (rare gas solids, metals, alkali and silver halides, etc) and defect processes, see Refs 6, 19-23. Moreover, it has been found to describe the parameters of the defect processes associated with the electric signals observed when applying uniaxial stress on ionic crystals, which are important for the explanation of the electric signals that precede earthquakes [28-30]. Thus, if $g^{act,x}$ and $g^{act,I}$ correspond to the activation Gibbs energies for the mixed crystal and the pure crystal (*I*), respectively we can write:

$$\frac{g^{act,x}}{g^{act,I}} = \frac{c^{act,x}}{c^{act,I}} \frac{B\Omega}{B_I\Omega_I} \qquad (10)$$

For crystals belonging to the same class, to a first approximation, we can assume that $c^{act,x}$ varies only slightly versus *x*, thus Eq.(10) gives

$$\frac{g^{act,x}}{g^{act,I}} = \frac{(1+x\lambda)^2}{1+x\mu\lambda} \qquad (11)$$

We now take into account that the electrical conductivity $\sigma$, for a single conduction process, can be written as [6]

$$\sigma \propto \exp\left(-\frac{g^{act}}{kT}\right) \qquad (12)$$

Therefore, by assuming that the preexponential factor in Eq.(12) does not vary significantly with composition, we conclude that the variation of the conductivity with the composition stems from the function $(1+x\lambda)^2/(1+x\mu\lambda)$ appearing in the right hand side of Eq.(11). The study of this function shows that it reaches a minimum value (by considering the assumption $\lambda > 0$) when the molar concentration $x$ of the constituent *II* takes the value

$$x_m = \frac{\mu - 2}{\lambda\mu} \qquad (13)$$

where we recall that the quantities $\mu, \lambda$ are explicitly obtained from the relations (5) and (7), respectively. Since the molar volumes of the end members are almost always known, the quantity $\lambda$ is easily deduced from Eq.(7) but the quantity $\mu$ is not usually experimentally accessible. A good approximation, however, of $\mu$ can be achieved by applying the aforementioned thermodynamical model as follows: By inserting Eq.(9) into the relation [2] $v^d = \left.\frac{dg^{act}}{dP}\right|_T$, we find $v^d = c^{act}\left[\left.\frac{dB}{dP}\right|_T - 1\right]\Omega$ and then applying the definition of $\kappa^d$ mentioned in Eq.(4), we finally get:

$$(\mu \equiv)\frac{\kappa^d}{\kappa_I} = 1 - \frac{B_I \left.\frac{d^2 B_I}{dP^2}\right|_T}{\left.\frac{dB_I}{dP}\right|_T - 1} \qquad (14)$$

In other words, the compressibility $\kappa^d$ of the defect volume can be approximately expressed through the elastic data of the pure component (*I*) (which is assumed here to have the higher concentration).

## 3. Application of Eq.(4) to the present case.

Let us now consider, for example, the mixed system AgBr-CdBr$_2$. The molar volumes of AgBr and CdBr$_2$ are $V_I$ =29 cm$^3$/mole and $V_{II}$ =52.43 cm$^3$/mole, respectively, which give –see Eq.(7)- $\lambda \approx 0.81$. We now proceed to the calculation of $\mu$ on the basis of Eq.(14) by using the elastic data under pressure obtained by Loje and Schuele [31]. They found that their data, as well as the early data obtained by Bridgman [32, 33], are well described if the expansion of the bulk modulus is carried out to second order, i.e.,

$$-\left(\frac{\partial P}{\partial \ln v}\right) = B(P) = B_0 + \left.\frac{dB_0}{dP}\right|_T P + \frac{1}{2}\left.\frac{d^2 B_0}{dP^2}\right|_T P^2 \qquad (15)$$

the investigation of which yields a second order Murnaghan equation (cf. The subscript "0" corresponds to values close to zero pressure). The resulting expression for the bulk modulus of AgBr was found to be [31]:

$$B(P) = 377.7 + 7.49 P - \frac{1}{2}(0.0287) P^2 \qquad (16)$$

where $B$ and $P$ are in kilobars, thus

$$\left.\frac{dB}{dP}\right|_T = 7.49 \quad \text{and} \quad \left.\frac{d^2 B}{dP^2}\right|_T = -0.0287 \qquad (17)$$

By inserting these values into Eq.(14) we find $\mu = \kappa^d/\kappa_I = 2.67$. The substitution of this value into Eq.(13) –after recalling that $\lambda$ =0.81 as mentioned above- we find $x_m \approx 30\%$. This may have an uncertainty of around ±15% if we consider the large experimental errors mainly involved in the way we extract the value of $\left.\frac{d^2 B}{dP^2}\right|_T$ from the measurements [34].

The calculated $x_m$ value is marked with an arrow in the horizontal axis of Figure 1(b), which depicts the measured values of the conductivity of the mixed system versus the composition for the AgBr-CdBr$_2$. The agreement between the calculated $x_m$ value and the experimental results is challenging, especially if we consider: (i) there exists an almost "flat" maximum of the $\sigma$-values versus $x$ (thus reflecting an experimental uncertainty in stating the exact experimental value of $x_m$) and (ii) the value of $x_m$ was calculated on the basis of a thermodynamical model by using solely the elastic data of the "pure" AgBr.

By the same token, a similar $x_m$ value is obtained from the aforementioned thermodynamical model for the system AgCl-CdCl$_2$, see Figure 1(a). This was expected upon considering that the "compression" curves $V$ vs $P$ deduced from the measurements in the "pure" crystals AgBr and AgCl (see Figures 3 and 4 of Ref. 31) exhibit a similar feature.

The following remark might be worthwhile to be added: Eq.(13), as we have shown above, provides the composition $x_m$ at which the conductivity of the mixed system attains its maximum value. This $x_m$ value must lie, of course, between zero and unity, i.e.,

$$0 \leq \frac{\mu - 2}{\lambda \mu} \leq 1$$

Since $\lambda$ is positive (by definition), we have $(\mu - 2)/\mu \geq 0$ and $(\mu - 2 - \lambda\mu)\lambda\mu \leq 0$. The first inequality gives

$$\mu \leq 0 \text{ or } \mu \geq 2, \tag{18}$$

whereas the second inequality leads to

$$(1 - 2/\mu) \leq \lambda \qquad (19)$$

We therefore conclude that systems with physical properties that do not satisfy these inequalities should not exhibit a maximum conductivity regardless of their composition. These inequalities are actually satisfied in the two systems investigated here, i.e., AgBr-CdBr$_2$ and AgCl-CdCl$_2$, thus explaining the appearance of a maximum when plotting $\sigma$ versus $x$.

A criticism against the procedure followed here could be raised on the following grounds: The present procedure is applicable only if we are justified to consider the system AgX-CdX$_2$ as a solid solution particularly at the temperature of our interest, i.e., below 650$^o$K. The borders between solid solutions based on silver halides and the two-phase-solid state regions have been confirmed in Ref. [12] by means of electrical conductivity measurements. A narrow solid solution (< 4.5 mol% AgBr) was also found to be formed on the CdBr$_2$ side (see Fig.3a of Ref. [12]). The solid solubility in both systems decreases upon decreasing the temperature and becomes negligible below about 400K. At the temperatures of isotherms taken into consideration here, however, the authors of Ref. [12] reported that there exist solid solutions based on AgCl and AgBr, thus allowing the application of our procedure.

**4. Conclusion**

Very recent measurements of the electrical conductivity of the mixed systems AgBr-CdBr$_2$ and AgCl-CdCl$_2$ have shown that the conductivity attains a maximum value for a concentration around 20 mol% of the cadmium halide. Here, we showed that the appearance of such a maximum can be explained if we consider a thermodynamical model that interconnects the Gibbs activation energy with the bulk properties just by taking into account the equation of state measured for the pure crystals (i.e., AgBr and AgCl) involved in the two mixed systems under consideration. In addition, general conditions based on the elastic data of the pure end members in a solid solution have been derived (i.e., the relations (18) and (19)), for the first time, which predict when a conductivity maximum at a certain concentration (given by Eq.(13) is expected to occur.

**References**

[1] S. Endou, Y. Michihiro, K. Itsuki, K. Nakamura, T. Ohno, Solid State Ionics 180 (2009) 488.

[2] B. Amrani, F. El Haj Hassan, M. Zoaeter, Physica B 396 (2007) 192.

[3] Y. Li, L. Zhang, T. Cui, Y. Ma, G. Zou, D.D. Klug, Phys. Rev. B 74 (2006) 054102.

[4] D.J. Wilson, A.A. Sokol, S.A. French, C.R.A. Catlow, Phys. Rev. B 77 (2008) 064115.

[5] D.J. Wilson, A.A. Sokol, S.A. French, C.R.A. Catlow, J. Phys. Condens. Matter. 16 (2004) S2827.

[6] P. Varotsos, K. Alexopoulos, Thermodynamics of Point Defects and their Relation with the Bulk Properties, North- Holland, Amsterdam (1986).

[7] I. Ebert, J. Teltow, Ann. Phys. 15 (1955) 268.

[8] A.P. Batra, L.M. Slifkin, J. Phys. Chem Solids 38 (1977) 687.

[9] P.A. Varotsos, K. Alexopoulos, J. Phys. Chem. Solids 39 (1978) 759.

[10] P. Varotsos, K. Alexopoulos, J. Phys. C: Solid State 12 (1979) L761.

[11] P. Varotsos, N. Sarlis, and E. Skordas, Acta Geophysica Polonica 50 (2002) 337.

[12] A. Górniak, A. Wojakowska, S. Plińska, E. Krzyźak, J. Therm. Anal. Cal. 96 (2009) 133.

[13] A. Wojakowska, S. Plińska, J. Josiak, E. Kundys, High Temp. High Press 30 (1998) 113.

[14] R. Blachnik, J.E. Alberts, Z. Anorg, Allg. Chem. 489 (1982) 161.

[15] A. Wojakowska, A. Górniak, A. Wojakowshi, High Temp. High Press 34 (2002) 349.

[16] D. Kostopoulos, P. Varotsos, S. Mourikis, Canadian J. Phys. 53 (1975) 1318.

[17] P. Varotsos, D. Miliotis, J. Phys. Chem. Solids 35 (1974) 927.

[18] P. Varotsos, Phys. Status Solidi B 90 (1978) 339.

[19] P. Varotsos, K. Alexopoulos, K. Nomicos, Phys. Status Solidi B 111 (1982) 581.

[20] P. Varotsos, Solid State Ionics 179 (2008) 438.

[21] P. Varotsos, K. Alexopoulos, Phys. Status Solidi A 47 (1978) K133.

[22] P. Varotsos, K. Alexopoulos, J. Phys. Chem. Solids 41 (1980) 443.

[23] P. Varotsos, K. Alexopoulos, J. Phys. Chem. Solids 42 (1981) 409.

[24] V. Katsika-Tsigourakou, A. Vassilikou-Dova, J. Appl. Phys. 103 (2008) 083552.

[25] C.M. Padma, C.K. Mahadevan, Mater. Manuf. Processes 22 (2007) 362.

[26] C.M. Padma, C.K. Mahadevan, Physica B 403 (2008) 1708.

[27] P. Varotsos, Phys. Status Solidi B 100 (1980) K133.

[28] P. Varotsos, K. Eftaxias, M. Lazaridou, K. Nomicos, N. Sarlis, N. Bogris, J. Makris, G. Antonopoulos and J. Kopanas, Acta Geophysica Polonica 44 (1996) 301.

[29] P. Varotsos, K. Eftaxias, M. Lazaridou, G. Antonopoulos, J. Makris, J. Poliyiannakis, Geophys. Res. Lett. 23 (1996) 1449.

[30] P. Varotsos, N. V. Sarlis, E. S. Skordas, S. Uyeda, M. Kamogawa, Proc Natl Acad Sci USA 108 (2011) 11361.

[31] K.F. Loje, D.E. Schuele, J. Phys. Chem. Solids 31 (1970) 2051.

[32] P.W. Bridgman, Proc. Am. Acad. Arts Sci. 74 (1940) 21.

[33] P.W. Bridgman, Proc. Am. Acad. Arts Sci. 76 (1945) 1.

[34] S.N. Vaidya, G.C. Kennedy, J. Phys. Chem. Solids 32 (1971) 951.

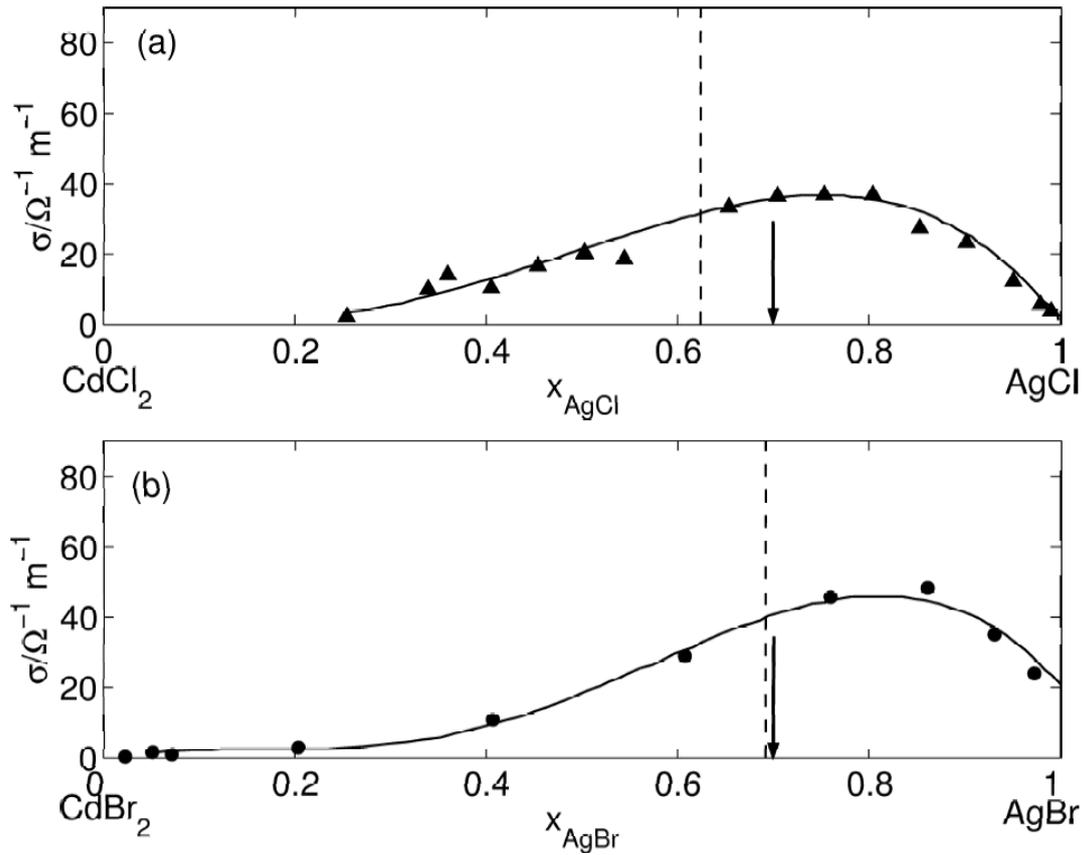

**Fig. 1** (a)-Conductivity for the system AgCl-CdCl$_2$ versus the concentration x of AgCl at 623 K. (b)- The same as in (a) but for the system AgBr-CdBr$_2$ at 648 K. The data are taken from Ref. [12]. The arrows in (a) and (b) mark the concentration at which the conductivity is predicted to become maximum from the thermodynamical model described in the text.